\documentclass{article}
\usepackage{spconf,amsmath,graphicx}

\usepackage{xcolor} 
\usepackage{hyperref} 
\usepackage{float} 
\usepackage{subcaption} 
\usepackage{booktabs} 
\usepackage{multirow} 
\usepackage{ragged2e}
\usepackage{amssymb, pifont} 
\usepackage{colortbl}

\title{Investigating on Incorporating Pretrained and Learnable Speaker Representations for Multi-Speaker Multi-Style Text-to-Speech}
\name{
    Chung-Ming Chien \quad
    Jheng-Hao Lin $^*$ \quad
    Chien-yu Huang $^*$ \quad
    Po-chun Hsu $^*$ \quad
    Hung-yi Lee \quad
    \thanks{$^*$ These authors contributed equally.}
}
\address{
    National Taiwan University\\
    College of Electrical Engineering and Computer Science\\
    \{r08922080, r08922049, r08921062,
    f07942095, hungyilee\}@ntu.edu.tw
}
\begin{document}
\ninept

\maketitle

\begin{abstract}
The few-shot multi-speaker multi-style voice cloning task is to synthesize utterances with voice and speaking style similar to a reference speaker given only a few reference samples.
In this work, we investigate different speaker representations and proposed to integrate pretrained and learnable speaker representations.
Among different types of embeddings, the embedding pretrained by voice conversion achieves the best performance.
The FastSpeech 2 model combined with both pretrained and learnable speaker representations shows great generalization ability on few-shot speakers and achieved 2nd place in the one-shot track of the ICASSP 2021 M2VoC challenge.
\end{abstract}
\begin{keywords}
    speaker representation, multi-speaker text-to-speech, few-shot
\end{keywords}

\vspace{-0.7mm}
\section{Introduction}
\label{sec:intro}
\vspace{-0.3mm}

Neural text-to-speech (TTS) has been proven to be capable of generating high-quality and human-like speech.
Previous research shows that the quality of the utterances synthesized by modern TTS models is already comparable with real human speech \cite{shen2018natural, ren2020fastspeech}.
However, when considering to generate speech of multiple speakers with a single model, the performance of such multi-speaker TTS models is still inferior to the single-speaker ones, especially when there is not enough high-quality data for any single speaker.
On the other hand, the speech generated by TTS models usually tends to be neutral and less expressive compared to real human speech.
As a result, improving the model's ability to model speaker and style variation has become an important topic in TTS research.

To push the frontier of TTS technology, ICASSP 2021 M2VoC challenge \cite{xie2021M2VoC} aims at addressing the problem of few-shot multi-speaker multi-style voice cloning.
In the challenge, a TTS system is required to generate speech with speaker identity and style similar to a few reference speech samples.
The TTS system must accurately model the variation of speech of different speakers with different speaking styles while maintaining the synthesized audio quality to achieve good results in the objective and subjective evaluations.
Under the few-shot setting, one major challenge of this task is to discover speaker and style information from limited references efficiently.
Previous research about multi-speaker TTS typically uses a speaker representation to control the synthesized utterance's speaker identity.
This speaker representation can be jointly learned with the TTS model in the form of an embedding table \cite{ping2018deep, sercan2018neural, chen2019sample,park2019multi} or a speaker encoder \cite{sercan2018neural, chen2019sample, cai2020from}, or can be transferred from another pretrained model for speaker information extraction \cite{jia2018transfer, cooper2020zero}.
While to control the style of synthesized speech, global style token (GST) \cite{wang2018style} is widely used to enable utterance-level style transfer.
Some also proposed to use an auxiliary style classification task \cite{wu2019end, li2020controllable} to disentangle style information from phonetic information in the utterances.
Since speaker and style information is usually entangled in the training data, it is also possible to learn a latent representation to jointly model the speaker and style information \cite{wang2018style, hsu2019hierarchical}.

In this work, we apply pretrained and jointly-optimized speaker representations to multi-speaker TTS models.
Two different TTS frameworks, Tacotron 2 \cite{shen2018natural} and FastSpeech 2 \cite{ren2020fastspeech}, are studied.
It is shown that with the jointly-optimized speaker representations only, the TTS models do not generalize well on the few-shot speakers.
We also demonstrate that using different pretraining tasks results in significant performance differences.
By combining both the pretrained and the learnable speaker representations, our experiments show that the audio quality and the speaker similarity of the synthesized speech improve significantly.
The synthesized samples are available online \footnote{Audio samples: \href{https://github.com/ming024/FastSpeech2/tree/M2VoC}{https://github.com/ming024/FastSpeech2/tree/M2VoC}}.
The results with the FastSpeech 2 TTS framework achieved 2nd place in the one-shot track of the ICASSP 2021 M2VoC challenge.

\vspace{-0.7mm}
\section{Speaker Representations}
\label{sec:speaker_representations}
\vspace{-0.7mm}

In the ICASSP 2021 M2VoC challenge, every speaker is related to a certain speaking style and vice versa, so there is no need to disentangle speaker and style information.
As a result, we use the speaker representations to control both the speaker identity and speaking style of synthesized utterances.
The pretrained and jointly-optimized speaker representations used in this work are introduced here.

\vspace{-0.7mm}
\subsection{Pretrained speaker representation}
\label{ssec:pretrained_speaker_embedding}
\vspace{-0.7mm}

Some pretraining tasks efficiently extract speaker information from speech.
For example, a speaker classification model predicts an input utterance's speaker identity and thus is widely used for speaker information extraction.
D-vector and x-vector are representative ones, and have been used in TTS \cite{jia2018transfer, cooper2020zero}.
On the other hand, since TTS is a generative task, it may be beneficial to apply the speaker representation pretrained on another generative task, such as voice conversion (VC), but this idea has never been investigated before.  
Three different pretrained speaker representations are discussed in this work.

\vspace{-0.7mm}
\subsubsection{D-vector}
\label{sssec:d_vector}
\vspace{-0.7mm}

D-vector \cite{wan2018generalized} is a speaker representation generated by a neural network pretrained on a speaker classification task, and is widely used in speaker verification systems.
Previous works \cite{jia2018transfer} have reported the effectiveness of combining a well-trained d-vector model with a TTS model to achieve high-quality multi-speaker TTS.

\vspace{-0.7mm}
\subsubsection{X-vector}
\label{sssec:x_vector}
\vspace{-0.7mm}
Similarly, x-vector \cite{snyder2018x} is also trained on a classification task, and was once considered a SOTA approach for speaker verification.

\vspace{-0.7mm}
\subsubsection{VC representation}
\label{sssec:vc_representation}
\vspace{-0.7mm}
One-shot VC models typically take a source utterance and a target utterance as inputs to synthesize an utterance with phonetic content similar to the source utterance while sounds like spoken by the target speaker. 
Hence, the module for target speaker information extraction is suitable for the pretrained speaker encoder in the multi-speaker TTS system.
As a one-shot VC system, AdaIN-VC \cite{chou2019one} can perform VC even when only one utterance of the target speaker is given, and was shown to have good generalizability to unseen speakers.
We take the target speaker encoder, which is used to extract speaker information from the target utterance, as a pretrained speaker encoder for the multi-speaker TTS systems.

\vspace{-0.7mm}
\subsection{Jointly-optimized speaker representation}
\label{ssec:jointly_optimized_speaker_representation}
\vspace{-0.7mm}
We can also optimize the speaker representations end-to-end while training the TTS model.
Some previous works suggest using multi-task training, such as jointly optimizing the TTS framework and an auxiliary classification task \cite{wu2019end, li2020controllable}.
However, it is rather complicated to balance multiple losses, so this work focused on the approaches without auxiliary losses.
Two different speaker representations are studied, the looked-up embedding and the GST.

\vspace{-0.7mm}
\subsubsection{Looked-up embedding}
\label{sssec:look_up_embedding}
\vspace{-0.7mm}
It is naive to apply a speaker embedding table to multi-speaker TTS.
By assigning an ID to each speaker from a predefined set of speakers, the representation vector for each speaker can be looked up from a trainable embedding table.
One limitation of this approach is that it cannot generalize to speakers unseen during the training stage.

\vspace{-0.7mm}
\subsubsection{Global style token (GST)}
\label{sssec:gst}
\vspace{-0.7mm}
We follow the setting of GST-Tacotron \cite{wang2018style} to use a convolutional RNN reference encoder and a style-token layer with multi-head attention to extract a sentence-level representation from a reference utterance.
The GST module is jointly trained with the entire TTS model without introducing additional loss terms. The GST representations are considered to contain paralinguistic information of the utterance, including speaker identity and speaking style, etc.

\vspace{-0.7mm}
\section{Workflow}
\label{sec:workflow}
\vspace{-0.7mm}

\begin{figure*}[tb]
    \centering
    \includegraphics[width=148mm]{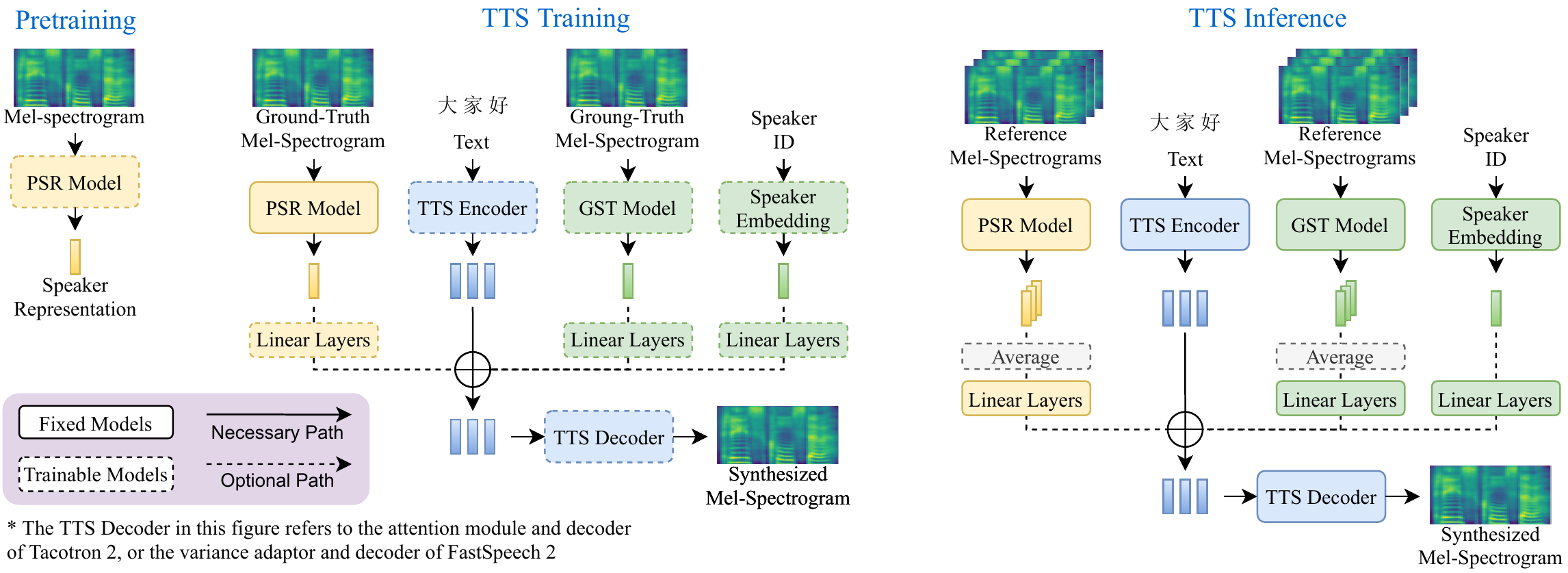}
    \vspace{-2mm}
    \caption{
        The workflow for our TTS framework. The PSR model in the figure stands for the pretrained speaker representation model.
    }
    \label{fig:workflow}
    \vspace{-5.5mm}
\end{figure*}

There are three stages in our workflow, including speaker representation pretraining, TTS training, and TTS inference.
An overall view is shown in Fig. \ref{fig:workflow}.

\vspace{-0.7mm}
\subsection{Pretraining}
\label{ssec:pretraining}
\vspace{-0.7mm}
The models for pretrained speaker representation extraction, including d-vector, x-vector, and the AdaIN-VC model, are first trained to extract an utterance-level 128-dimensional representation from a mel-spectrogram input.
The pretrained models will not be finetuned in the following stages.

\vspace{-0.7mm}
\subsection{TTS training}
\label{ssec:tts_training}
\vspace{-0.7mm}
In this stage, the TTS model and the jointly-optimized speaker representations (including the looked-up embedding table and the GST) are end-to-end trained.
Different combinations of the pretrained and learnable speaker representations are explored in our experiments, such as using the VC representation and the looked-up embedding in a single TTS model.
The speech representations provide speaker information to the TTS model to perform multi-speaker TTS.

Two TTS frameworks, Tacotron 2 and FastSpeech 2, are studied in this work.
Tacotron 2 is a sequence-to-sequence model composed of an encoder, a decoder, and an attention module.
The encoder takes phoneme inputs to generate a hidden feature sequence, which the decoder then attends.
The decoder autoregressively predicts the mel-spectrogram with the aid of the attention module.

On the other hand, as a non-autoregressive model, FastSpeech 2 comprises a Transformer encoder, a Transformer decoder, and the variance adaptor in between.
The encoder again generates the phoneme-level hidden features from the phoneme inputs.
The variance adaptor consists of a pitch predictor, an energy predictor, and a duration predictor.
The pitch and energy predictors take the phoneme-level hidden features as inputs to predict the prosody features, which are then combined with the hidden features to control the prosody of synthesized speech.
Afterward, since the duration of each phoneme is not uniform in speech, the duration predictor predicts the length of each phoneme and then expands the phoneme-level features according to the predicted lengths.
In the end, the decoder generates mel-spectrograms from the expanded features \footnote{In the original FastSpeech 2 \cite{ren2020fastspeech}, the pitch and energy predictors are placed after the duration predictor to predict frame-level prosody features. However, in our implementation, we find that performing phoneme-level prosody prediction before expanding the phoneme-level features helps to improve the synthesized audio quality.}.

To control the speaker identity, the pretrained and jointly-optimized speaker representations serve as additional inputs to the TTS model, as shown in Fig. \ref{fig:workflow}.
For Tacotron 2, the utterance-level speaker representation is projected to match the size of the output of the encoder with a two-layer linear network.
The projected representation is then expanded through time and added to the output of the encoder before applying the attention mechanism.
For FastSpeech 2, the projected speaker representation is also expanded and then summed with the output of the encoder, so the pitch, energy, and length prediction is conditioned on the speaker representation.
Both the pretrained and jointly-optimized speaker representations (except the looked-up embedding, which is mapped from the speaker ID) are extracted from the ground-truth mel-spectrogram at training time.

\vspace{-0.7mm}
\subsection{TTS inference}
\label{ssec:tts_inference}
\vspace{-0.7mm}
At inference time, to imitate a particular speaker's voice, we extract a speaker representation from each utterance spoken by the speaker and compute an averaged representation.
We also tried to use the representation of a randomly selected utterance of the speaker, but an averaged representation makes the quality of the synthesized utterances more stable.

\vspace{-0.7mm}
\section{Experiments}
\label{sec:experiments}
\vspace{-0.7mm}

\vspace{-0.7mm}
\subsection{Task description}
\label{ssec:task_description}
\vspace{-0.7mm}
The official datasets of the ICASSP 2021 M2VoC challenge consist of 4 subsets, the MST (including MST-AIShell and MST-Originbeat), TSV, TST, and TT subsets.
The former three subsets include audio recordings with transcriptions, while the text-only TT subset contains the texts used for testing-time synthesis.
There are two tracks in the challenge, the few-shot track (track 1) and the one-shot track (track 2).
In track 1, the TTS systems are required to synthesize utterances of the few-shot speakers in the TST dataset given 100 reference utterances of each speaker, while in track 2, only 5 utterances are given.
The synthesized utterances are evaluated on the audio quality, speaker similarity, and style similarity.

\vspace{-0.7mm}
\subsection{Training setup}
\label{ssec:training_setup}
\vspace{-0.7mm}
All of the recordings, including that of the few-shot speakers, were used to train the TTS model. 
No external data was used.
The audio files were resampled to 22050 kHz and used to extract 80-dimensional mel-spectrograms.
The mel-spectrograms were used as the input of the speaker representation modules and the TTS models' training targets.
A WaveNet vocoder \cite{oord2016wavenet}\footnote{WaveNet: \href{https://github.com/r9y9/wavenet_vocoder}{https://github.com/r9y9/wavenet\_vocoder}} trained on the MST-Originbeat, TSV, and TST subsets was used to convert the mel-spectrograms back to the raw waveform.

For the speaker representation pretraining, we used the open-source implementations of d-vector\footnote{d-vector: \href{https://github.com/yistLin/dvector}{https://github.com/yistLin/dvector}}, x-vector\footnote{x-vector:\href{https://github.com/manojpamk/pytorch_xvectors}{https://github.com/manojpamk/pytorch\_xvectors}}, and AdaIN-VC \footnote{AdaIN-VC: \href{https://github.com/cyhuang-tw/AdaIN-VC}{https://github.com/cyhuang-tw/AdaIN-VC}}.
These models were trained on the MST and TSV datasets from scratch.
Each model extracted an utterance-level 128-dimensional speaker representation given a mel-spectrogram input.
After the pretraining, the pretrained speaker representation models were fixed, and the TTS models combined with different speaker representations were trained end-to-end.
The Tacotron 2 \footnote{Tacotron 2: \href{https://github.com/BogiHsu/Tacotron2-PyTorch}{https://github.com/BogiHsu/Tacotron2-PyTorch}} models were trained for 200k steps, while the FastSpeech 2 \footnote{FastSpeech 2 (for final submission to the ICASSP 2021 M2VoC challenge): \href{https://github.com/ming024/FastSpeech2/tree/M2VoC}{https://github.com/ming024/FastSpeech2/tree/M2VoC}} models were trained for 500k steps, both with batch size 16.

FastSpeech 2 requires the ground-truth duration, pitch, and energy for each phoneme to train the variance adaptor.
We used Montreal Forced Aligner \cite{mcauliffe2017montreal} trained on the MST, TSV, and TST datasets to align the speech with the text transcriptions to get the duration of each phoneme.
The frame-level pitch values were extracted with PyWorldVocoder \footnote{PyWorldVocoder: \href{https://github.com/JeremyCCHsu/Python-Wrapper-for-World-Vocoder}{ https://github.com/JeremyCCHsu/Python-Wrapper-for-World-Vocoder}}, while the L2 norm of each short-time-Fourier-transform frame was used as the energy descriptor.
The pitch and energy of each phoneme were obtained by computing the phoneme-wise average of the frame-level values. 

\vspace{-0.7mm}
\subsection{Objective evaluation}
\label{ssec:objective_evaluation}
\vspace{-0.7mm}
A speaker verification (SV) task was used to evaluate the speaker similarity between the synthesized speech and the real speech.
We used a third-party pretrained speaker encoder \footnote{Resemblyzer: \href{https://github.com/resemble-ai/Resemblyzer}{https://github.com/resemble-ai/Resemblyzer}} to extract speaker embeddings from utterances.
We computed the average of the speaker embeddings of every utterance of each speaker as the enrollment embedding of that speaker.
A synthesized utterance passed the SV system as long as the cosine similarity between the embedding of the utterance and the enrollment embedding of that speaker exceeded a threshold, which was obtained by computing the equal error rate over the whole dataset.
The equal error rate was 5.90 \% throughout the MST, TSV, and TST datasets.

The results of the SV test with different TTS models are shown in Table \ref{tab:objective}.
In slot (a) and (b), only one speaker representation was used for each TTS model.
It can be seen the VC representation consistently outperforms all other representations in both the Tacotron 2 and FastSpeech 2 experiments, while the performance of the jointly-optimized speaker representations is not good under the few-shot setting. 
Despite that the results of FastSpeech 2 are better than Tacotron 2, we should note that the Tacotron 2 models were only trained for 200k steps, which may not be enough for full convergence.

To study the effect of using multiple speaker representations in one TTS model, we combined the VC representation with other representations for FastSpeech 2.
The results are shown in slot (c) of Table \ref{tab:objective}.
We do not see any obvious difference between the results of Track 1, while in Track 2 it is observed that combining the VC representation with x-vector or the looked-up embedding is better than any single speaker representation. 
The best result is achieved by combining the VC representation and the looked-up embedding, resulting in 93.7 \% speaker verification accuracy, when only 5 reference utterances of each speaker are given.

\begin{table}
    \setlength{\tabcolsep}{2.35pt} 
    \centering
    \caption{
        The results of the objective speaker verification tests for TTS models with different types of speaker representations.
    }
    \vspace{-1.5mm}
    \label{tab:objective}
    \begin{tabular}{cccccccc}
        \toprule
        
        \multirow{3}{*}{\textbf{Model}} & \multicolumn{5}{c}{\textbf{Speaker Representation}} & \multicolumn{2}{c}{\textbf{Results}} \\
        & \multicolumn{3}{c}{Pretrained} & \multicolumn{2}{c}{Learnable} & \multicolumn{2}{c}{SV Accuracy} \\
        \cmidrule(lr){2-4}
        \cmidrule(lr){5-6}
        \cmidrule(lr){7-8}
        & d-vec & x-vec & VC & embed & GST & Track 1 & Track 2 \\
        \midrule
        
        \multirow{5}{*}{\textbf{(a) Tacotron 2}}
        & \checkmark & & & & & .772 & .367 \\
        & & \checkmark & & & & .785 & .377 \\
        & & & \checkmark & & & .942 & .727 \\
        & & & & \checkmark & & .630 & .703 \\
        & & & & & \checkmark & .102 & .050 \\
        \midrule

        \multirow{5}{*}{\textbf{(b) FastSpeech2}}
        & \checkmark & & & & & .977 & .323 \\
        & & \checkmark & & & & .973 & .623 \\
        & & & \checkmark & & & .980 & .837 \\
        & & & & \checkmark & & .988 & .490 \\
        & & & & & \checkmark & .778 & .340 \\
        \midrule

        \multirow{6}{*}{\textbf{(c) FastSpeech2}}
        & \checkmark & & \checkmark & & & .978 & .747 \\
        & & \checkmark & \checkmark & & & \textbf{.992} & .860 \\
        & & & \checkmark & \checkmark & & .983 & \textbf{.937} \\
        & & & \checkmark & & \checkmark & .982 & .783\\
        \cmidrule(lr){2-8}
        & \cellcolor{gray!15} & \cellcolor{gray!15} & \cellcolor{gray!15} \checkmark & \cellcolor{gray!15} \checkmark & \cellcolor{gray!15} \checkmark & \cellcolor{gray!15} .988 & \cellcolor{gray!15} .897 \\
        & \checkmark & \checkmark & \checkmark & \checkmark & \checkmark & .990 & .887 \\
        \bottomrule
        
    \end{tabular}
    \justify
    \vspace{-2mm}
    \footnotesize $^\ast$ The colored row is the model used for the final submission to the ICASSP 2021 M2VoC challenge. Due to the time limitation, we did not submit our best model.
    \vspace{-1.3mm}
\end{table}

\vspace{-0.7mm}
\subsection{Subjective evaluation}
\label{ssec:subjective_evaluation}
\vspace{-0.7mm}
Subjective evaluations were conducted to further confirm the results.
We used two 5-scale mean opinion score (MOS) tests to evaluate the quality and speaker similarity of the synthesized utterances.
Three FastSpeech 2 models with one speaker representation and one FastSpeech 2 model with combined VC representation and looked-up embedding were tested.
In the quality MOS test, the native speaker listeners were given one synthesized utterance of the Track-2 speakers and were asked to give a score between 1 to 5 points for the audio quality.
In the similarity MOS test, one real utterance and one synthesized utterance were given to the subjects, and they gave a score for the speaker similarity between the two utterances.
Each model received at least 300 scores in each test. Table \ref{tab:objective} shows the results.

Similar to the objective evaluation, it is observed that the speaker similarity of the model combining the pretrained and jointly-optimized speaker representations is superior to all other models.
As for the audio quality, the model with combined speaker representations is almost as good as using VC representations or looked-up embeddings only, and is much better than x-vector.
The results of the subjective evaluations indicate that combining pretrained and learnable speaker representations effectively improves the performance of multi-speaker TTS models.

\begin{table}
    \setlength{\tabcolsep}{4.1pt} 
    \centering
    \caption{
        The results of the MOS tests of audio quality and speaker similarity for the Track-2 speakers. FastSpeech 2 models with different combinations of speaker representations are compared.
    }
    \vspace{-1.5mm}
    \label{tab:subjective}
    \begin{tabular}{ccccc}
        \toprule
        
        \multirow{2}{*}{\textbf{Model}} & \multicolumn{4}{c}{\textbf{Speaker Representation}} \\
        \cmidrule{2-5}
        & x-vec & VC & Embed & VC+Embed \\
        \midrule
        MOS\scriptsize{quality} & 3.47 $\pm$ .13 & 3.61 $\pm$ .13 & \textbf{3.65} $\pm$ .13 & 3.55 $\pm$ .12 \\
        MOS\scriptsize{similarity} & 3.25 $\pm$ .13 & 3.19 $\pm$ .14 & 3.27 $\pm$ .13 & \textbf{3.38} $\pm$ .14 \\
        \bottomrule
        
    \end{tabular}
    \justify
    \vspace{-7.3mm}
\end{table}

\vspace{-0.7mm}
\subsection{Analysis}
\label{ssec:analysis}
\begin{figure}[tb]
    \centering
    \vspace{-2.5mm}
    \includegraphics[width=90mm]{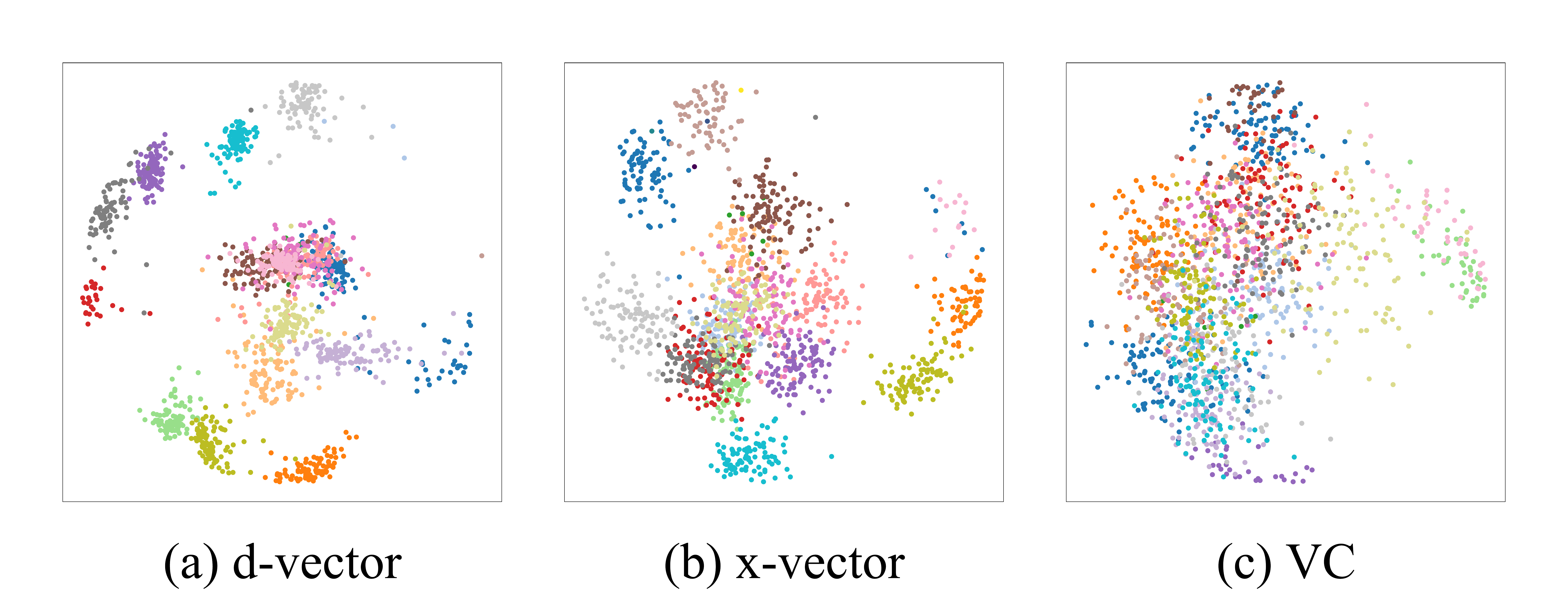}
    \vspace{-5mm}
    \caption{
        Visualization of different speaker representations. Different colors stand for different speaker identities.
    }
    \label{fig:pca}
    \vspace{-3mm}
\end{figure}

\vspace{-0.7mm}
To get more insight into the speaker representations, we visualize the speaker representations in Fig. \ref{fig:pca}.
100 utterances of 20 randomly sampled speakers are plotted.
The 128-dimensional vectors are reduced to 2-dimensional with principal component analysis.
GSTs are not included here since they are jointly trained with the TTS model, and their behavior changes in different training scenarios.

Compared with d-vector and x-vector, the latent distribution of the VC representations is more continuous.
It is because both d-vector and x-vector are pretrained on a discriminative task.
The models only aim to separate the utterances of different speakers, so they may ignore the information that does not help discriminate different speakers, which may be harmful to the TTS task.
An ideal speaker representation for TTS models should model every characteristic of the speakers, but it is not necessary to have clear boundaries between speakers.
VC representation contains the information needed for speech generation. 
This may be the possible reason that the VC representation outperforms other pretrained representations in the objective and subjective evaluations.
However, this hypothesis still needs to be proved with more supportive experiment results.

\vspace{-0.7mm}
\subsection{Official subjective evaluation}
\label{ssec:official_subjective_evaluation}
\begin{figure}[tb]
    \centering
    \vspace{-1.2mm}
    \includegraphics[width=90mm]{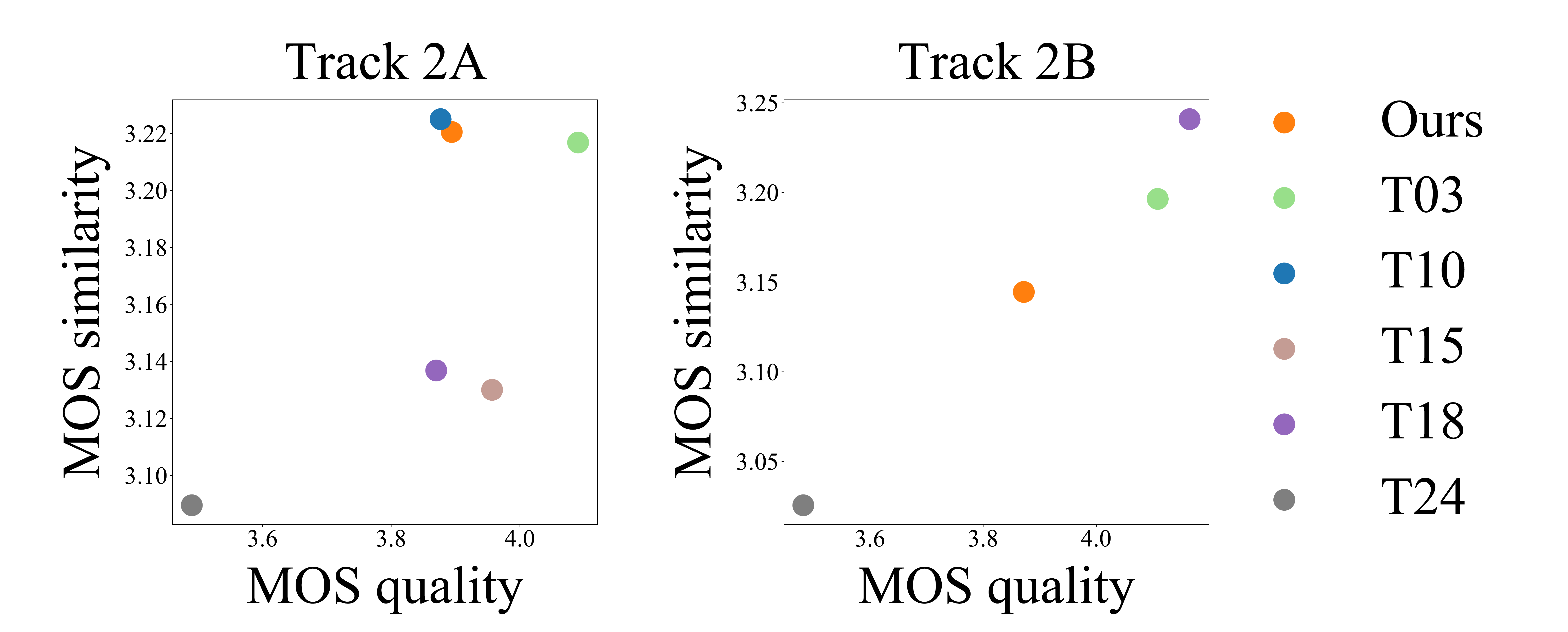}
    \vspace{-5.1mm}
    \caption{
        The official subjective evaluation results of Track 2.
    }
    \label{fig:official}
    \vspace{-5.5mm}
\end{figure}

\vspace{-0.7mm}
The official evaluation of the ICASSP 2021 M2VoC challenge includes three subjective MOSs (quality, speaker similarity, style similarity) for Track 1 and two subjective MOSs (quality, speaker similarity) for Track 2.
Our approach shows competitive results with other participants, especially in the one-shot track.
In track 2A, where external training data is not allowed, our results ranked 2nd out of 17 teams.
Even when external training data is allowed in track 2B (but we did not use any), our results are only inferior to 2 other teams out of 19 teams.
The scores of several top-ranked teams of Track 2 are shown in Fig. \ref{fig:official}

\vspace{-0.7mm}
\section{Conclusion}
\label{sec:conclusion}
\vspace{-0.7mm}
In this work, we apply different types of speaker representations to the multi-speaker multi-style voice cloning task.
We show that for TTS applications, the speaker representations pretrained on generative tasks may be better than those pretrained on discriminative tasks. 
We propose to combine pretrained speaker representations with jointly-optimized speaker representations, which yields better results than using any single representation.
Results of the objective and subjective evaluations confirm the effectiveness of the proposed approach, which points out a simple yet effective direction for few-shot multi-speaker TTS.

\vspace{-0.7mm}
\section{Acknowledgement}
\label{sec:acknowledgement}
\vspace{-0.7mm}
We thank to National Center for 
High-performance Computing (NCHC) of Taiwan for providing computational and storage resources.

\pagebreak

\bibliographystyle{IEEEbib}
\bibliography{main}

\end{document}